\RequirePackage{fix-cm}
\documentclass[twocolumn]{svjour3}          
\smartqed  
\usepackage{graphicx}
 \usepackage{url}
\usepackage[comma,authoryear]{natbib}
 \journalname{Environmental Geochemistry and Health}
\begin{document}

\title{
Chronic Kidney Disease of Unknown aetiology (CKDu)
and multiple-ion interactions in drinking water.
}


\titlerunning{CKDu and multiple-ions.}        
\authorrunning{Dharma-wardana} 

\author{ M.W.C. Dharma-wardana       
}

\institute{M.W.C. Dharma-wardana        \at
National Research Council. Ottawa, Canada, K1A 0R6, and
          Universit\'{e} de Montreal,  Canada, H3C 3J7l.   \\
              \email{chandre.dharma-wardana@nrc-cnrc.gc.ca}     
}

\date{Received: }

\maketitle

\begin{abstract}
Recent experimental work on the nephrotoxicity of contaminants in drinking water
using laboratory mice, motivated by the need to understand the origin of chronic
 kidney disease of unknown aetiology is examined within 
our understanding of the hydration of  ions and proteins. Qualitative considerations
based on Hofmeister-type action  of these ions, as well as  quantitative
electrochemical models for the  Gibbs free-energy change for ion-pair formation
are used to explain why
Cd$^{2+}$ in the presence of F$^-$ and water hardness due to Mg$^{2+}$ ions (but  not
 Ca$^{2+}$) can be expected to be more nephrotoxic, while AsO$_3^{3-}$ in the presence
 of F$^-$ and hardness may be expected to be less nephrotoxic. The analysis is
 applied to a variety of ionic  species typically found in  water 
 to predict their likely combined electro-chemical action. These results
clarify the origins of chronic kidney disease in the
north-central province of Sri Lanka. The conclusion is further strengthened by a
study of the dietary load of Cd and As, where the dietary loads are found
to be safe, especially when the mitigating effects of micronutrient ionic forms of
 Zn and Se,
as well as corrections for bio-availability are taken in to account. The resulting
aetiological picture supports the views that  F$^-$, 
Cd$^{2+}$ (to a lesser extent), and Mg$^{2+}$ ions  found in stagnant  
household well water  act together with enhanced toxicity, becoming
the most likely causative factor of the disease.
Similar incidence of CKDu found in other  tropical climates may
have similar geological origins. 
\end{abstract}
\section{Introduction}
\label{intro}
Many disadvantaged tropical  communities are facing a new type of kidney disease
 even though  the recognized causes (e.g., diabetes,
hypertension, etc.) of kidney diseases are absent. It is generally called `Chronic Kidney Disease of Unknown
 aetiology' abbreviated to CKDu or CKDU. Urine-albumin excretion above standard
 thresholds is usually the early clinical sign of CKDu. 
Increasing incidence of CKDu among adult rural
populations in  Sri Lanka as reported in \cite{WHO1} and by
 \cite{WHO2}, while \cite{DVReddy2013} have reported on India. Similarly,
 El Salvador
\citep{Trabanino05},  China \citep{BoLin14}, South-Asia \cite{ckdu-Pakistan}
and the  global distribution~\citep{GlobalDimension2015}  have come  under
scrutiny during the last decade. 
 We deal with experiments on mice that were undertaken as a step 
towards understanding the nature of  CKDu in Sri Lanka; but the
 results have wider implications~\citep{SynergyBandara2017}.

The North central province (NCP) of Sri Lanka was home to  the ancient
Anuradhapura civilization ($\sim5$ century BCE to 10 century CE). It is a part
of the so-called  ``dry zone" (DZ) of Sri Lanka, with a rainfall $\sim 1750$ to 1000
mm. per annum~\citep{Agri-dept} and largely overlaps with the region affected
by CKDu.  Because the illness is most prevalent in disadvantaged 
agricultural communities, it
was natural to look for agrochemical and environmental 
explanations for the illness. Many of the proposed  causes of CKDu
were based on the toxicity of pesticides, or heavy-metals found in
fertilizers. Others looked for a geo-genic origin, e.g.,
fluoride~\citep{Dissanayake2005,IllepAl}, or  origins in occupational health (e.g., dehydration in a
dry climate). However, the concentrations of heavy-metal toxins or pesticides found in the
water or soil in the affected area were negligible or inconclusive
~\citep{WHO2,NanayakkaraS14} in comparison to the maximum allowed levels (MALs)
stipulated by WHO and other regulatory bodies. In contrast, urine samples and
biopsies~\citep{WHO2,Paranagama2012, LevineWanig2016} of patients showed levels of
 Cd, As, and
Pb well above the expected amounts. Given that most patients were in a
relatively advanced state of the disease when diagnosed, these high levels
merely indicated the bio-accumulation of toxins which occurs when kidneys
become impaired. Phyto-accumulation of Cd and As were also found, and above the MALs in
staples like rice and most types of vegetables consumed by residents
in  the affected areas {\it as well as} in CKDu-free areas~\citep{Premarathne2006,
WHO2, Premarathne2011}. 
 The first comprehensive study of CKDu by~\cite{WHO2} in Sri Lanka was 
sponsored by the World Health organization (WHO)
and the National Science Foundation (NSF) of Sri Lanka. The study tentatively
suggested that Cd may
be the cause of CKDu, although no firm conclusion was made.
Whether the toxin load from food is relevant to CKDu
in Sri Lanka or not will also be discussed (sec.~\ref{diet.sec}).

In the following we review evidence which suggest that sub-MAL quantities of
ions like fluoride become nephrotoxic in the presence of hard water, and also
 increase the nephrotoxicity of cadmium but not that of arsenic. A qualitative discussion
 using  Hofmeister-type protein-denaturing concepts~\citep{SalisNinham2014, Collins1997}  
 given in 
~\cite{Dharma-wardana2015} is replaced by a more quantitative electrochemical model.
The presence of Cd, As and F in food,
and the co-presence of counter agents like Se and Zn, the dietary load,  and how the MALs
given by regulatory organizations should be interpreted will also be
examined.

\section{Toxins in drinking water.}
In a previous publication by ~\cite{Dharma-wardana2015} two
probable  scenarios for the origin of the disease were presented, and 
 drinking water was selected as the likely mediating agent of the disease.
The first scenario was based on the observation of ~\cite{Jayasekera2012} that many
 of the affected communities were adjacent to
river basins of large irrigation projects that could serve as conduits for fertilizer
runoff from plantations. Judging from World-Bank data~\citep{WorldBankFert},
El Salvadore, a country affected by CKDu,  is found to use less
than 1/26 of the agrochemicals/hectare compared to a high-end user like  New Zealand
(2013 data).
Similarly, Sri Lanka uses on the average only  about  1/5 to 1/10 the  agrochemicals per
hectare used by New Zealand (1836 kg/hectare in 2013) which is free of
chronic kidney disease of unknown aetiology. 
This anti-correlation of agrochemical usage with disease
is not sufficient to rule out agrochemicals as a cause since  
amounts that exceed the mean is used in the plantation areas of Sri Lanka's 
central
hills. Hence the hypothesis of agrochemical runoff into the NCP via rivers and
possible accumulation needed further examination. However, recent work by 
\cite{Diyabalanage2016} has shown the following:\\
 {\scriptsize ``...the trace metal
concentrations in the Mahaweli (river) upper catchment were detected in the order of
Fe $>$ Cu $>$  Zn $>$  Se $>$  Cr $>$  Mn $>$  As $>$  Ni $>$  Co $>$  Mo. Remarkably
high levels of Ca, Cr, Co, Ni, Cu, As, and Se were observed in the Mahaweli Basin
compared to other study rivers. Considerably high levels of Cr, Mn, Fe, Co, Ni, Cu,
Zn, As, and Se were found in upstream tributaries of the Mahaweli River...Cd ... often
attributed to the etiology of unknown chronic kidney diseases ... is much lower than ... reported levels''.}\\
 Thus
traditional ``heavy metal'' toxins like Cd, As, Pb and Hg  are not flowing into the
endemic areas along ``irrigation rivers''. 
Furthermore, Cd, As
 and other
``heavy metals'' found in the rice grown in the rain fed {\it wet zone} exceed the
 amounts in the
 dry zone by 40-60\% ~\citep{Diyaba-Rice-2016,Meharg2013} and so one cannot attribute any
 clear role in the aetiology of CKDU to the irrigation rivers that feed the dry zone. 

Chemical analysis of the soil and water~\citep{WHO2} and
by \cite{NanayakkaraS14, LevineWanig2016} in
the affected areas shows that heavy metal concentrations are generally well below
 the maximum
allowed levels (MALs) stipulated by regulatory bodies. There is, however
a  runoff of PO$_4^{3-}$ ions into the NCP as seen from Table 1 of
~\cite{Dharma-wardana2015}. But there are no MALs stipulated for phosphate in
drinking water, even though it is a  strongly Hofmeister-active ion. The second
scenario that was presented by ~\cite{Dharma-wardana2015} was the  action of Panabokke's 
redox processes in stagnant shallow
wells of the NPC. This gradually increases the ionicity of the well waters. The  origin of
 kidney disease was linked to  water of high ionicity
damaging the membranes in the tubules and other components of the kidney via the
Hofmeister mechanism. The suspicion that users of drinking water
from household wells could be at risk was  in the literature by
2011~\citep{KamaniCd11,IllepAl}. Early work had showen
 that  well waters in endemic areas
have high electrical conductivities. 
Detailed studies of the wells  and isotopic evidence for the lack
of connectivity of these wells to the main water table of the region have been presented
more recently by ~\cite{Manthrithilake2016}. The confluence of fluoride, hardwater and
cadmium in the endemic areas has also been established by \cite{WasanaWaterQ-CdF2016}.
Mouse studies using NCP water from three locations and a control were
presented by \cite{ThammitiyaNCPwater2012}, providing  evidence of nephrotoxicity.
Hence the
studies of  ~\cite{SynergyBandara2017} on experimental mice
 fed with water containing
controlled amounts of fluoride, heavy
metals and hardness become very topical and clarify the
aetiology of the disease. 
\begin{table*} 
\caption{A summary of  the results of tissue damage of experimental mice by
various combinations of ions in drinking water~\citep{SynergyBandara2017}), as
reflected in the weight change  in grams per mouse. Maximum Allowed Limits (MALs)
indicate the safe limits for each substance, according to WHO drinking water
standards used in Ref.~\cite{SynergyBandara2017}, i.e., MALs for F=1.5 mg/l, 
Cd=3 $\mu$g/l, As=10 $\mu$g/l, Pb=10 $\mu$g/l,
secondary standard Al=200 $\mu$g/l, Hardness: CaCO$_3$=200 mg/l, and MgCO$_3$=185
mg/l .}
\center
\begin{tabular}{p{0.5 cm} p{1.5 cm} p{1.5cm} p{1.5 cm}p{2.8 cm} p{3.0 cm }p{1.2cm} p{2.2 cm}}
\hline

no.

$\downarrow$ & ion$_1$ &does &  ion$_2$ & dose  & hardness &dose   & Weight-change g/mouse\\                       
\hline\\

1    & Cd &$<<$MAL     & F &$<<$MAL     & (Ca,Mg)CO$_3$  & $<<$MAL  & +9  \\
2    & Cd &MAL         & F &MAL         & (Ca,Mg)CO$_3$  & MAL      & -3  \\                                                        
3    & Cd &2MAL        & F &MAL         & (Ca,Mg)CO$_3$  & 2MAL     & -7 \\                                              
4    & Cd &MAL         & --& --         & (Ca,Mg)CO$_3$  & MAL      & -4 \\
5    & Cd &MAL         &  F &MAL         &   --         & --       & -5  \\
6    & --& --         &  F &MAL         & (Ca,Mg)CO$_3$  & MAL      & -3 \\
7    & Cd &2MAL        & -- &--          &  --          & --       & -3 \\
7F   & --             &-- & F &0.03MAL-6.7MAL &  --  & --        &--\\  
8    & As & 1.5 MAL    & F &2MAL        & (Ca,Mg)CO$_3$  & MAL      & -5 \\
9    & Pb & 1.5 MAL    & F &2MAL        & (Ca,Mg)CO$_3$  & MAL      & -3 \\
10   & Al & 1.0 MAL    & F &2MAL        & (Ca,Mg)CO$_3$  & MAL      & -1 \\
11   & As & 1.0 MAL    & -- &   --          & --             & --       & -5 \\
12   & Al,Cd,As&MAL    & F &MAL         & Ca,Mg)CO$_3$   & MAL      & -3 \\
\hline
\end{tabular}
\label{ion-damage.tab}
\end{table*} 

\section{Experimental results on the nephrotoxicity of combinations
 of ions in drinking water.} 
The
experimental results of Wasana  et al relevant to our study are summarized in
Table~\ref{ion-damage.tab}.  \cite{SynergyBandara2017} give details of
kidney-tissue damage which include (i) interstitial fibrosis, (ii) mononuclear cell
interstitial inflammation, (iii) tubular atrophy, (iv) apoptosis, (v) cell
degeneration, (vi) cell necrosis,  (vii) Glomeruler lesions, (viii) glomerulosclerosis,
and (ix) hyperemia and hemorrhage. In order to have a simple quantity characterizing
the impact of the insult on the organism, we have selected the average weight loss (or
gain) per mouse during the experiment as a simple index and presented it in the
tabulation. The weight loss corresponds fairly well with the detailed damage profiles
given by ~\cite{SynergyBandara2017}. 

The results of ~\cite{SynergyBandara2017} establish quite definitively that
although fluoride is not nephrotoxic even at $\sim$6.7 times the MAL 
(i.e. 10 mg of F$^-$ per litre of soft water; see row 7F of
 table~\ref{ion-damage.tab}), it becomes
nephrotoxic in the presence of hard water (see row 6). It is clearly more nephrotoxic
than even cadmium in soft water. In fact row 7 shows that it takes twice the MAL of Cd to
reach the same level of kidney  damage and weight loss as fluoride at the MAL in hard
water. The WHO and other MAL specifications on fluoride are based on fluorosis rather
than nephrotoxicity and there has been less recognition of the nephrotoxicity of
fluoride. In fact, row 7F of the table shows lack of renal toxicity at this concentration. 
The MALs quoted by other regulatory bodies can be somewhat different. For instance, the 
US Food and Drugs Administration gives an MAL of 
5 $\mu$g/l for Cd~\citep{ATSDR2013}, while the MAL
for fluoride has been lowered to 0.7 mg/l of fluoride~\citep{Reuters2011}. The 
US Environment Protection Agency has used 4 mg/l as the MAL for fluoride. The need to
set MALs with reference to the hardness and simultaneous presence of fluoride in
drinking water becomes clear from the work of  Wasana  et al.
\subsection{Relevance of the results to human nephrotoxicity.}
The experimental
results should be a strong indication of how human kidney tissue will also respond to
these toxic insults, if the data can be related to a human model.
 The life expectancy of  male Sprague-Dawley mice (used in the
experiment) is about three years, and the duration of the experiment ($\sim$150
days) corresponds to about 14\%  
of the organism's life time, i.e., in human terms, about 10 years. This in fact
corresponds to the likely time scales for the onset of CKDu. On the other hand,
as pointed out by Vijaya  ~\cite{VijayaKumarPrCom2017}, dosage levels are about
three times higher than what occurs in the context of CKDU. Supplementary material (Table S 3)
of Wasana et al state that the mice imbibed around 30 ml/week or about 5 ml/day of
water. 
Given  a weight
of $\simeq$30 g/mouse, this corresponds (linear scaling) to a human intake of
 10 l/day of water with the stated concentrations of ions. At a more
typical human intake of 3 l/day, the corresponding fluoride concentrations would be 
0.45 mg/l instead of 1.5 mg/l, and 0.9 $\mu$g/l of cadmium instead of 3$\mu$g/l.
The data of  \cite{LevineWanig2016} for the composition of water in dug wells
give a maximum of 0.168 $\mu$g/l of Cd, and 2.05 mg/l of fluoride (but no 
standard deviations are given), indicating the daily intakes one may expect.
 If we use a linear model and scale
the observed toxicities (in terms of weight loss per mouse) to an intake of
 about 1/3 the amount of water, then the 
weight changes for rows 2-6 of  table~\ref{ion-damage.tab} become roughly -1, -2.6, -1.3, -1.6, and
-1 g/mouse, still showing significant toxicities above the likely noise threshold of the
experiment. 

\section {Hofmeister-type  action of ions in water.}
\label{ionss-neph.sect}
The chronic action of corrosive ions on  kidney tissues can be qualitatively
 understood
 from the interaction of ions and water with protein substrates. 
Biological properties of these membranes involve targeted folding and 
unfolding of proteins which pump ions and water along biochemically
defined gradients. Thus the  mechanisms of both beneficence and toxicity
are similar at the molecular level, with the correct ions or incorrect ions 
docking at target sites in proteins and creating good 
or disruptive behaviour.

Ions themselves carry a sheath of water molecules, i.e., a
``solvation shell'' of relatively tightly bound ``captive water'', while at least
a  loose second  solvation shell exists. In strongly charged ions like Al$^{3+}$ even a third
sheath is relatively tightly bound, so that the effective  radius $r$
 of the solvated ion in water is increased compared to the intrinsic ionic radius $r_i$
in {\it vacuuo}. The 1st hydration shell adds $\sim$ 1.4 \AA$\,$ to the intrinsic ionic
radius~\citep{Shannon1976}. Hence the effective electric field $Z/(\epsilon r)$
 of the ion in water with charge $Z$ is significantly decreased. 
Here the dielectric-screening constant  $\epsilon$ of the
solvated water around the ion is $\simeq 1.5-2$, i.e.,
much less than that of bulk water where $\epsilon\simeq 80$. 
 In Table~\ref{radii.tab} we
give ion radii, nominal numbers of solvated water molecules (solvation number $n_s$)
 and  Gibbs free
 energies of hydration for some ions of interest. 
The mean solvation numbers $n_s$ may
range from 4-8 or more for common cations, and 2-8 for some anions; The $n_s$ values
 are sometimes optimized to non-integer values $n_s^*$ for use in physico-chemical models
 in fitting to experimental data~\citep{Marcus87}. 


Composite ions like AsO$_3^{3-}$, or CO$_3^{2-}$, have a  behaviour different to simple
 ions like Na$^+$
as their
`surfaces' exposed towards the water medium are negatively charged oxygen ions. 
The latter interact loosely through hydrogen bonding and  `disrupt' the tetrahedrally
hydrogen-bonded structure of water. The effective radius of the anion (AsO$_3)^{3-}$
 is about $\sim$2.5 \AA, and is determined by its trigonal geometry, its As-O bond
 length of 1.77\AA,
 and the large oxygen atoms.
Proteins also from
a hydration layer at their charged surfaces where electrostatic effects
and  hydrogen bonding come into play.
 The water attached to the protein
membranes, glycans in glycoproteins etc., play a vital role in ion channels and
 fenestrae in the endothelium,
and across the basement membrane (a network of proteins) of the kidney.
While  acute toxicity acts directly via specific reactions
targeting specific chemical pathways usually leading to oxidative stress,
 chronic toxicity works by more subtle
pathways, one of which is the change in the hydration sheaths of protein layers 
or intermediates like the Kidney-injury molecule (KIM-1), leading to local
restructuring of protein layers. 
 A qualitative description
of modifications of water structure, both in bulk and on protein layers is afforded
by the concepts of ``kosmotropes''(order-makers) and
 ``chaotropes''(disorder-maker)~\citep{SalisNinham2014, Collins1997}. 
Ions like Mg$^{2+}$ are kosmotropes while AsO$_3^{3-}$ with its
trigonal structure has a chaotropic effect on the tetrahedral water structure. However,
while such discussions can be suggestive, they are hard to quantify.
Modern methods for relating structure to activity~\citep{Nilar2012}
may be used for toxicity studies, but this remains a largely
unexplored area. In this study
we assume that the association of ions in water to form ion pairs leads to a disruption of
how these ions are docked or interact with  protein layers, and 
 create  the possibility of toxic injury.  
%
%
\begin{table*}
\caption{Ionic radii, coordination number and the hydration free energy.}
\label{radii.tab}
\center
\begin{tabular}{p{2.5 cm}p{0.9 cm}p{0.9 cm}p{1.2 cm}p{1.2 cm}p{1.3 cm}p{0.9 cm}p{0.9 cm}p{0.9 cm}p{0.9 cm}p{0.9 cm}}
\hline

  ion $\to$

item$\downarrow$   & Al$^{3+}$ & Ca$^{2+}$ &  Cd$^{2+}$ & F$^-$    & Fe$^{3+}$  & Mg$^{2+}$  &Pb$^{2+}$  &CO$_3^{2-}$ &AsO$_3^{3-}$& NO$_3^-$\\       
\hline\\ 
radius/\AA$\,
$(Marcus)          &0.53       & 1.0-1.1   & 0.8-0.95 & 1.3-1.33 &  0.65      & 0.66-0.72 & 1.19      & 1.8        & 2.4-3.0    & 1.79\\
sol. number $n_s$  & 6         & 6-8       & 4-6      & 4-6      &  16.6      &  4-6      & 6.0       & 3-4        & 3-4        & 2   \\

$-\Delta G$
 kJ/mole           & 4525      & 1505      &  1755    &  465     & 4265       & 1830      & 1425      &1315        & 2608       &300  \\
\hline
\end{tabular}
\label{ion-G.tab}
\end{table*}

%
\section{Electrochemical approach to
synergies and antagonisms in the nephrotoxicity of mixtures of ions in water.}
\label{syn-antag.sub}
While Hofmeister concepts like Kosmotropy, Chaotropy
and  qualitative measures like
the  ``principle of matching water affinities''
provide qualitative guides, they are difficult to quantify. 
Instead we use a semi-empirical electrochemical approach.

In clarifying  the synergistic  action of ions, we  consider:
(a) two types of ions, e.g., Mg$^{2+}$ and F$^-$, or Cd$^{2+}$ and F$^-$ ions,
 being present in water together
and  found to modify their toxicity; (b) two types of ions found to act independently,
e.g., the case of Ca$^{2+}$ and AsO$_3^{3-}$ ions,
and (c) other cases involving multiple ions. The formation of ion-pairs
or their absence is proposed as the model for elucidating
the  observed toxicities, and the model seems to correlate well
with the experimental observations.

Pair formation between the cation $C$ and anion $A$ occurs
if the Gibbs free energy of the hydrated ion pair is less than
the sum of their respective  individual hydration free energies.
\begin{equation}
\label{pair.eq1}
\Delta G_p(CA) < \Delta G(C)+\Delta G(A).
\end{equation}
This is the most basic expression of co-solvant/solute effects,
while various reformulations, e.g.,  based on preferential binding concepts etc., are
used in the study of protein-salt interactions~\citep{Ben-Naim1992,Smith2004}.
%
%
%
Here we consider only ion-ion interactions; 
interactions with proteins are not directly invoked in the current formulation.
 Gibbs free energies for ion-pairing is calculated using the electrochemical model of
~\cite{Marcus87}. More advanced solvation models, e.g., ~\cite{Rasiah1998},
as well as microscopic
quantum mechanical calculations using density functional simulations 
(e.g., \cite{XiaJinDFT-AlF2011})  will
be left for future work.
 
The pairing free energy $\Delta G_p$ between an anion A and a cation C contains
 an energy term and
an entropy terms. These contributions to pairing of hydrated ions
involve explicit changes in the hydration of the ions.
\begin{eqnarray}
\label{pair-eq2}
C(H_2O)_{n_c}+A(H_2O)_{n_a} &\leftrightarrow& 
CA(H_2O)_{n_{ca}}+\\
& &(n_c+n_a-n_{ca})H_2O
\end{eqnarray}
The  Coulomb energy of pairing for a ``contact pair'' CA is of the form
\begin{equation}
\label{Coulomb-pair.eq}
\Delta E=Z_cZ_a e^2/\{\epsilon(r_c+r_a)\}
\end{equation}
where $Z_a$, and  $r_a$  are the valance and radius 
respectively of the anion, and
similarly for the cation where the subscript `$c$' is used.
The valency of anions is taken as negative.
 the dielectric constant  $\epsilon=1$. Our objective
is the Gibbs free energy change $\Delta G_p^0$ per mole
under standard conditions of pressure and temperature (STP) for the ion
 pairing proposed in eq.~\ref{pair-eq2}.
\begin{figure}[b]
\label{ion-pair.fig}
\begin{center}
\vspace{0.1 cm}
\includegraphics[width=0.75\linewidth]{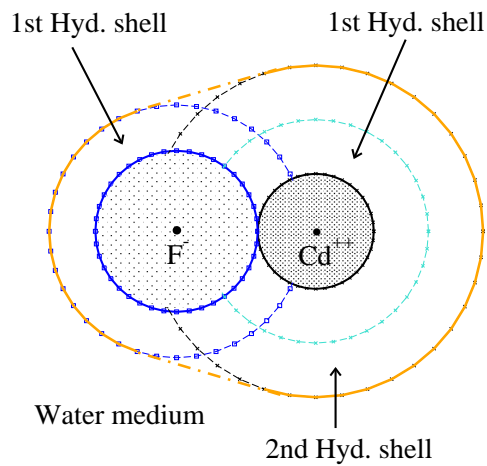}
\caption{(On line colour) A schematic diagram of the (Cd-F)$^+$ ion-pair
 together with the
 hydration shells of the  Cd$^{2+}$ ion and the F$^-$ ion prior to pairing.
The divalent Cd$^{2+}$ ion holds tightly two hydration shells, while the
monovalent F$^-$ holds only a single hydration shell. The water outside
the joint hydration shell of the  pair `sees' an ion of effective valance
 $Z_p=1$. The reduction in solvation energy  on pairing
is offset by the Coulomb energy when persistent ion pairs are formed.}   
\end{center}
\end{figure}
The $r_c,r_a$ values refer to the smallest coordination radii of the cation
and the anion when contact pairing occurs.
The new ion pair of valance $Z_p=Z_c+Z_a$ has its
own hydration free energy. We use the semi-empirical fit formulae  given
 by ~\cite{Marcus87} to evaluate $\Delta G_p$.

We consider an isolated ion in water
 of valance $Z$, effective radius $R=r+\Delta r$,
 where $\Delta r$ is the radial increment due to
 hydration.  $\Delta G$ of hydration of a single ion placed in water
 is the sum of the following terms.
\begin{eqnarray}
\label{e1.eqn}
\Delta G_{1 }&=&(N_{av}e^2/2)Z^2\{1-1/\epsilon'\}\{\Delta r/(\Delta r R)\}\\
\label{e2.eqn}
\Delta G_{2 } &=& (N_{av}e^2/2)Z^2\{1-1/\epsilon\}\{1/(\Delta r R)\}\\
\label{e0.eqn}
\Delta G_{0 }  &=&41-87(R/\mbox{nm})+1200(r/\mbox{nm})^2;\; \mbox{kg/mole}\\
\label{e-as.eqn}
\Delta G_{as}  &=&1200Z^3 r; \; \mbox{kg/mole}
\end{eqnarray}
The hydration energy evaluated using these formulae are close to the
experimental values, but may be in error by about $\pm$ 10\%.
The constant $N_{av}$ is
the Avogadro number, and Gaussian units are employed. 
The Gibbs free energy change due to the strongly-held hydration shell
is given by $\Delta G_1$ and invokes $\epsilon'$, the dielectric constant of
 the strongly-held water. A value of $\epsilon'\simeq 1.79$ is 
 used in the formulation of Marcus.
 $\Delta G_2$ is the contribution from the weakly held
hydration shells spreading into the water medium, and here $\epsilon\simeq76.9$.
The last two terms, $\Delta G_0, \Delta G_{as}$ are semi-empirical
and are given in practical units of kJ/mole, with the radii
in nm.  The fact that
the water molecules orient differently at a cation from those at an anion (where 
the protons point to the anion) is accounted for via $G_{as}$. The term $\Delta G_0$
is a semi-empirical neutral-solute correction that
occurs in the Born-Haber solvation cycle. 

The above model can be applied to an ion pair $p$ made up of the Cation + Anion.
 In addition to the
 Coulomb energy of the pair, hydration
energy terms similar to Eqs.~\ref{e1.eqn}-\ref{e-as.eqn} are needed;  $r_p=r_c+r_a$,
and  $Z_p=Z_c+Z_a$ are the radius and valance of the ion pair. The change in the Gibbs
hydration free energy on pairing is the sum of the following terms.
\begin{eqnarray}
\label{e1p.eqn}
\Delta G_{1p}&=& \{\Delta G_1(C)+\Delta G_1(A)\}\frac{V_h(p)}{V_h(A)+V_h(C)} \\
\label{e2p.eqn}
\Delta G_{2p} &=& (N_{av}e^2Z_p^2/2)\{1-1/\epsilon\}\{1/(\Delta r_{p} r_{p})\}\\
\label{e0p.eqn}
\Delta G_{0 }  &=&41-87(r_{p}/\mbox{nm})+1200(r_{p}/\mbox{nm})^2\\
\label{e-asy-p.eqn}
\Delta G_{as}^p &=& 1200Z_p^3 r_{p}. 
\end{eqnarray}
Eq.~\ref{e1p.eqn} contains the volumes of the hydration sheath of the anion, cation and
the pair, viz., $V_h(A),V_h(C),V_h(p)$ respectively. These are easily evaluated
from the geometry of the ion pair displayed in Fig.~1. 
The volume fraction
$V_h(p)/\{V_h(A)+V_h(C)\}$ is of the order of 0.4-0.7 for common ion pairs. In evaluating
the contribution for hydration beyond the captive shell, i.e., $\Delta G_{2p}$, and for
the asymmetry and neutral corrections, we assume that for $r>r_p$ the ion-pair is effectively
like a single ion of charge $Z_p$. The increment $\Delta r_p$ in $r_p$ from solvation 
is evaluated from the following equations with $r$/(nm).
\begin{eqnarray}
\Delta r&=&(0.3542-0.5587r)/(1+152.3r^2),\; Z_p=\pm 1, \\
\Delta r&=&(0.3779 -0.4750r)/(1+95.06r^2), \; Z_p=\pm 2, \\
\Delta r&=&(0.4126-0.5745r)/(1+71.01r^2),\; Z_p=\pm 3.
\end{eqnarray}
These  equations reproduce the $\Delta r$ of individual
cations and anions used by ~\cite{Marcus87}, Table 1.

The resulting pairing Gibbs free energy changes per mole, $\Delta G_p^0$ are given in
Table~\ref{pair-G.tab}. We have not included Cl$^-$ and Na$^+$ ions, as these are
 monovalent ions with relatively large radii with no tendency to associate.
The results for $\Delta G_p^0$ and $\Delta G_c^0, G_a^0 $  can now be used to calculate
 pair concentrations given the concentrations of reactants in any
experiment, using the relation
$G=G^0+RT\log_e\{\mbox{activity}\}$. The  activity may be taken to be approximately
 equal to the molar concentration for dilute solutions.
Further chemical speciation calculations (e.g, using ~\cite{MINTEQ} and such codes)
 are in fact
 not necessary for our purposes
where we are dealing with laboratory water used in the experiment of
\citep{SynergyBandara2017} at a pH of $\simeq$7.
Furthermore, standard codes may not 
adequately treat the relevant ion-pairing processes
due to the lack of accurate experimental equilibrium-constant data. 
 If we consider a binary pair like (Al-F)$^{2+}$,
it will also form more complex aggregates like AlF$_2^+$ etc.,
but the higher complexes are less important and the formation of
the binary complex (usually the dominant form) is the key to
 understanding the changed biophysics of nephrotoxicity.
Hence we directly use table~\ref{pair-G.tab} to determine
when pairing is likely to be relevant.

\begin{table}
\caption{Estimated Gibbs free energies (kJ/mole) $\Delta G_p^0$ per mole
for pair formation between some ions of interest.}
\center
\begin{tabular}{p{1.75 cm}p{1.0 cm}p{1.0 cm}p{1.0 cm}p{1.0 cm}}
\hline\\
$\;\;\;\;\;$  anion

cation$\downarrow$     &  F$^-$  & CO$_3^{2-}$ & AsO$_3^{3-}$ & PO$_4^{3-}$\\       
\hline\\
Cd$^{2+}$              & -13.7   & 352         & 1715         & 1625   \\
Ca$^{2+}$              & 159     & 534         & 1851         & 1769   \\
Mg$^{2+}$              & -545    &-130         & 1214         & 1205   \\
Al$^{3+}$              & -5782   & -4343       &-3996         & -3089  \\
Pb$^{2+}$              & 2852    & 3600        & 4948         & 4918   \\
Zn$^{2+}$              & -276    & 107         & 1382         & 1401   \\
Fe$^{3+}$              &-2588    &-2002        &-1175         &-936    \\
\hline \\
\end{tabular}
\label{pair-G.tab}
\end{table} 

\subsection{Nephrotoxicity in the presence of two types of ions.}
\label{syn-sub}
While Cd$^{2+}$ and AsO$_3^{3-}$ are well established as nephrotoxins when present
above their MALs, F$^-$ is not universally recognized as a nephrotoxin. However,
we treat it as a nephrotoxin under suitable conditions, as established from even
 the work of ~\cite{SynergyBandara2017}.
Furthermore, the effect of other ions commonly found in drinking water on the
 nephrotoxicity of 
Cd$^{2+}$, AsO$_3^{3-}$, and F$^-$ is not understood and we examine this in the light of
cation-anion interactions. There is no anion-anion or cation-cation pair formation
as the Coulomb interaction energy is positive for like charges. We also include in our
study the transition-metal ion Fe$^{3+}$ as its behaviour is typical of such
ions which have small ionic radii and high charge. They readily form ion aggregates
which are, however, usually insoluble. But binary pairs like (FeF)$^{2+}$ at low
concentrations  remain in solution, and can contribute to nephrotoxicity
although this aspect has received little or no attention. 
Panabokke's redox mechanisms that may operate in dug wells~\citep{Dharma-wardana2015}
emphasized iron and manganese (see also \cite{LevineWanig2016}),
 while the work of \cite{Diyabalanage2016} emphasized the
dominant presence of iron and copper in the Mahaweli-river water. We do not make a
detailed study of transition-metal effects, but present the data for Fe$^{3+}$ as being
typical of such ions.

\subsubsection{{\normalfont Cd}$^{2+}$ and one anion species present together.}
The first-row entries in Table~\ref{pair-G.tab} show that only the two ions 
Cd$^{2+}$ and F$^-$ form a
stable ion pair (CdF)$^+$ which has a radius of about 0.2 nm. The others,
including Cd$^{2+}$ and AsO$_3^{3-}$ have a positive free energy of formation
and hence will not associate together. Hence we consider only the (CdF)$^+$ ion pair.

The $\Delta G^0$ values for
the ion pair and for the individual ions Cd$^{2+}$, F$^-$ can be used to show that
 all the fluoride ions
used in the experiment of Wasana et al~\citep{SynergyBandara2017}, row 5, will convert to
  (CdF)$^+$ ion pairs with no water molecules between
them (contact pairs), but with a single common solvation sheath of water molecules
 (Fig.~1).
The  (CdF)$^+$ ion pair will appear as a ``sodium-like'' ion
and approach protein  sites that attract sodium ions,
and break apart to reveal  toxic cadmium and  fluoride 
ions - two bullets wrapped into one!

Thus  the toxicity of Cd, and fluoride are both enhanced by their
mutual presence. This is also observed from the experimental data of
Table~\ref{ion-damage.tab}. While cadmium alone at even twice the  MAL in soft water
produces a weight change  of -3 g/mouse (row 7) , cadmium jointly with fluoride 
(both at their MAL concentrations, row 5) produced a weight change of -5 g/mouse. 
Hence the joint toxicity of Cd$^{2+}$ and F$^-$  is
 very strong and equals  about 3 times that of cadmium alone at its MAL  in soft water
 (assuming a quasi-linear dependency of the toxicity on the dose for this limited range).

\subsubsection{{\normalfont Ca}$^{2+}$ and one anion species present together.}
Table~\ref{pair-G.tab}, row 2 shows that there is {\it no ion-pair formation} 
with Ca$^{2+}$ ions and hence the type of synergistic action shown in Wasana et al
cannot be caused by ion-pairing due to calcium ions in hard water.
 The chronic toxicity caused
 by high levels of arsenic in well waters has led to public health emergencies
 in Bangladesh,
Taiwan, and many other countries. Hence it was
natural to consider the possibility that the CKDu epidemic in the NCP of Sri Lanka may be 
connected with arsenic toxicity. 

However, to explain the geographic localization of the
disease to the NCP region, a number of workers proposed that 
 arsenic together with the hard water typical of the region was responsible
for the disease~\citep{Jayasumana11,jayasumana13} and that Ca$^{2+}$ ions conferred
an enhanced toxicity to AsO$_3^{3-}$ making even sub-MAL amounts of arsenic
nephrotoxic. However, the Gibbs free energies given in Table~\ref{pair-G.tab} 
for pairing, and in  Table~\ref{ion-G.tab} for Ca$^{2+}$ and AsO$_3^{3-}$ do not
support such a possibility within an ion-pair approach. In a subsequent revised version
 of this theory ~\citep{JayasumanaGly2015,TaskForceGly} it was
%
hypothesized  that Ca$^{2+}$ ions, AsO$_3^{3-}$, together with glyphosate 
mediated an enhanced toxicity to drinking water. Gibbs free energy calculations
 for Ca$^{2+}$ ions, AsO$_3^{3-}$,
 together with glyphosate inclusive of its hydration sheath also turn out to
 produce no stable intermediates; 
instead, Ca$^{2+}$ ions combine with glyphosate to form an insoluble product as has been
 known experimentally for decades~\citep{SmithRay88}. Since insoluble materials
precipitate out, they cannot be agents in any toxicity linked to drinking water.
\subsubsection{{\normalfont Mg}$^{2+}$  and one anion present together.}
Row 3 of the table of free energy changes predicts that magnesium ions pair
 with fluoride ions to form
(MgF)$^+$ ions. At the concentrations of Mg$^{2+}$ and F$^-$ specified by the
experiment of ~\cite{SynergyBandara2017}, calculations using the $\Delta G_p^0$
values from the Tables~\ref{ion-G.tab} and \ref{pair-G.tab} show that all the fluoride ions
are completely converted to the form  (MgF)$^+$.
The (MgF)$^+$ ion-pair looks like a Na$^+$ ion as
it  approaches a protein surface if the distance of approach
is significantly larger than $r_p\sim 0.2$ nm. At closer range
the Mg$^{2+}$ and F$^-$ will break apart under
the electric fields of the ionic centers and H-bonding sites of the protein.
The magnesium and fluoride ions are  delivered to the wrong sites, e.g., possibly
docking  sites in epithalial Na-channels, disrupting their reactive capacity and structure.
The ATPases that play a role in the sodium pump act via transport or exchange of monovalent
(H$^+$, Na$^+$, K$^+$) and various divalent ions~\citep{Na-Pump2007}. The divalent
 Mg$^{2+}$ will also be delivered to the wrong sites.
Hence fluoride ions in the presence of magnesium hardness may be expected to be
more nephortoxic than fluoride ions alone. This is in agreement with
the experimental data of row 7F and row 6, Table~\ref{ion-damage.tab}, where the
the experimental mice suffer a weight loss of 3 g/mouse.\\

It should be noted that Mg$^{2+}$ ions do not form ion pairs with AsO$_3^{3-}$, and hence
magnesium cannot be invoked to propose that CKDu in the NCP of Sri Lanka
is caused by AsO$_3^{3-}$ toxicity enhanced by this type of (i.e., magnesium) hardness. 
\subsubsection{{\normalfont Al}$^{3+}$ and one anion species  present together.}
The results from Table~\ref{pair-G.tab} clearly shows that  Al$^{+3}$ readily forms  ion
pairs with practically all the  anions. Aluminum in various forms is abundant in
the environment. However, Al$^{+3}$  strongly interacts not only with other anions, but
also  with colloidal matter, soil etc., found in water and precipitate out of
drinking water. Aluminum salts are used as flocculants in water treatment.
Thus small concentrations of aluminum ions are ubiquitous in drinking water,
with a non-mandatory MAL of 200 $\mu$g/l.
Healthy individuals flush out the ubiquitous but low-concentrations of
aluminum ions that enter their bodies, but a long-standing controversy about
the neurotoxicity of aluminum exists, especially when used as pharmaceutical
 adjuvants~\citep{Al-Shaw2011}.
 
While aluminum by itself  has not been seriously implicated  as
 a nephrotoxin relevant to CKDU,
it has  been proposed that fluoride ions may react with aluminum ions from
cooking utensils and contribute to CKDU by ~\cite{IllepAl} and
followed up by \cite{Amara-Dharmaguna2014}.
 Our results in Table~\ref{pair-G.tab}
show that F$^-$ will associate with Al$^{3+}$ in water to form
 (AlF)$^{2+}$ ions. Since aluminum has no specific bodily function 
the (AlF)$^{2+}$ ion pair  is not likely to be more toxic than F$^-$ itself.
 Furthermore, since Al$^{3+}$ 
pairs with most anions (and even with cations due to its amphoteric nature)
it can act to {\it reduce} toxicity by competing with other 
ions that form ion pairs e.g.,  (Mg$^{2+}$, F$^-$), as discussed below. 
A DFT-study of the exchange of the  solvated-water of Al$^{3+}$ with F$^-$
 ions has been
carried out by ~\cite{XiaJinDFT-AlF2011}.
%
Aluminum ions in the well water can help to dissolve fluoride ions from
the regolith and increase fluoride concentrations in stagnant wells.
%
\subsubsection{{\normalfont Pb}$^{2+}$  and one anion species  present together.}
The Gibbs free energy calculations in row 5 of Table~\ref{pair-G.tab} show that lead ions
 show
no tendency to associate with any of the anions listed. Hence, within the ion-pairing model,
 lead ions will have no
modifying effect on the toxicity of fluoride or arsenic contaminants in water (except for
the usual electrolytic effects  due to the modifications of activity coefficients). Since
 the toxic ion Cd$^{2+}$ is positively charged, there is of course no ion-pair formation
 between Ca$^{2+}$ and Pb$^{2+}$ either.
\subsubsection{{\normalfont Zn}$^{2+}$  and one anion species  present together.}
The Gibbs free energy calculations in row 6 of Table~\ref{pair-G.tab} show that Zn ions
  associate strongly with fluoride ions to form (ZnF)$^+$, giving strong competition
to cadmium. In fact, the Gibbs free energy change $\Delta G_p$ for (ZnF)$^+$ is
some 20 times larger (negative) and hence there will be virtually no (CdF)$^+$
ion pairs formed.  
If it is present in well water, Zn can increase the concentration of fluoride ions by
leaching it off the soil into the water. The ion-pair (ZnF)$^+$ is likely to be more toxic than
fluoride acting alone, just as is the case with magnesium ions.
\subsection{A transition-metal cation and one  anion species present together.}
A variety of transition-metal ions are found together  in water and
the situation proposed here, where the transition-metal cation being dominant
is unlikely in running water. However, such situations may arise in the stagnant
water in dug wells.  \cite{LevineWanig2016} report a maximum of 1.38 mg/l of
Fe, while the US-EPA MAL for Fe is 0.3 mg/l. However, many other ions
(F, Mn, Na, Pb) exceeded the stipulated MALs and well water is by no means a binary
ion system.  Nevertheless we study it as a a useful step.
 The values of $\Delta G_p$ for ferric ions show that any fluoride 
or arsenide ions would be converted into a binary pair with the ferric ion, until all the
fluoride ions, or arsenide ions (these being usually the ones with a lower molarity)
are exhausted. There is a strong possibility that the arsenide will be oxidized to
the less toxic arsenate form and  complex further to precipitate off the solution phase.
The (FeF)$^{2+}$ binary ion is likely to be more nephrotoxic than the fluoride ion
acting alone, but Fe$^{3+}$ is recognized as potent fluoride coagulant because of the
capacity of ferric ions to form such binary complexes and then coagulate with any organic colloidal matter present in the water, esp. under high pH 
conditions~\citep{KerslakeFeF1946}. Hence ferric ions may actually reduce the amount of fluoride ions present in natural water at elevated pH.
  
\subsection{Toxicity in the presence of multiple ions.}
Normal drinking water always contains a variety of ions, e.g., Al$^{3+} < $ 0.05-0.2 mg/l,
Ca$^{2+} <  $ 200 mg/l, Mg$^{2+} <  \; $150 mg/l, HCO$_3^{3-} < $ 200 mg/l,$\;$ etc.,  which 
have secondary MALs, while known toxins like  F$^- <$ 3 mg/l 
 Cd$^{2+} < 3 \mu$g/l, Pb$^{2+} <$, AsO$_3^{3-}$, Hg$^{2+}$ 
are  present at the level of a few $\mu$g/l and their MALs are primary standards, 
i.e., enforceable by law. The current primary standards are based
on the toxicity of the ion  acting alone. In general such an
assumption  is satisfactory because the presence of many types of ions leads to
a {\it reduction} in the toxicity of the multiple-ion system. In a 
multi-ion system there are
many pairing processes possible, most of which do not augment the toxicity, but compete
with the toxicity-enhancing pair processes. This may be cast into a more rigorous
statement based on the central-limit theorem, but the analysis will not be elaborated here.
Furthermore, in dealing with actual drinking water (rather than the example
of water used in the mouse experiment of \cite{SynergyBandara2017}), 
a chemical equilibrium model  for the calculation of chemical
 speciation, solubility equilibria, sorption, presence of suspended matter for natural waters
should be used  after ensuring that the needed free-energy data for the relevant ion species
are available in the data base of the software used.


If we consider the four component system containing cadmium, fluoride, aluminum
 and Mg$^{2+}$ from hard water (see the experiment of Wasana et al, row 12), it is
 effectively a  six-ion system, i.e, 
Cd$^{2+}$, F${^-}$, Al$^{3+}$, Mg$^{2+}$ 
 and  counter ions which may be  NO$_3^-$ for Cd$^{2+}$, and Na$^+$ for F$^-$,
or  any other simple  ions  considered to be nontoxic.
The ion pairs relevant to toxicity are
 (a) fluoride with aluminum
 (b) fluoride with magnesium, and (c) fluoride  and cadmium. 
Of these, the last pair, (Cd-F)$^+$, is strongly nephrotoxic with the toxicity enhanced,
 while (b) enhances the toxicity of fluoride. Hence, the formation
of (b) and (c) together competes with each other. The formation of the aluminum complex
leads to {\it no enhancement} of fluoride toxicity, and strongly competes with (b) and (c), 
with the highly favourable $\Delta G_p$ of Al$^{3+}$. The net effect, e.g., at typical
concentrations close to the MALs is to {\it reduce} the toxicity of the mixture due to
 the presence of
multiple ions. This is in fact observed in the experimental data
of  ~\cite{SynergyBandara2017}, (row 10 of Table~\ref{ion-damage.tab}), where
 the multi-ion mixture produces a weight loss of 3 g/mouse, where as
the (CdF)$^+$ pair alone would have produced a weight loss of 5 g/mouse(row 5). The decrease
in toxicity on adding competing Mg$^{2+}$ from hard water is seen in row 2 where the weight
 loss is 3 g/mouse.

The presence of many ions in natural drinking water leads to multiple interactions
most of which are non-toxic, and hence take away from the strength of the toxic 
interactions. This situation can be strongly modified in stagnant water in 
 wells which draw from disconnected underground regolith aquifers.  Specific
 salt species can concentrate due to a variety of geochemical and redox
 processes~\citep{Pana1,Dharma-wardana2015}, and also form concentration gradients
 in  such water wells~\citep{Manthrithilake2016}.


\section{Likely origin of trace metal toxins  and fluoride in drinking water.}
\label{origin.sec}
The presence of fluoride in the geological area endemic to CKDu is well
 accepted~\citep{WHO2}, while
its confluence with hard water has also been recognized. 
CKDu is prevalent among NCP residents who use certain shallow household wells with water
having a high electrical conductivity, but not among those who use water from
 reservoirs (known locally as ``weva''), irrigation canals and rivers.  These
 form an interconnected network of water bodies linking the CKDu stricken area with highly
agricultural areas nearby as well as  the central hills of the country. Analysis of
the flowing water  by a number of independent groups has revealed levels of
  Cd$^{2+}$ below the MALs~\citep{Diyabalanage2016, Jayasinghe-RO-2015, NanayakkaraS14, WHO2}. 
Nevertheless, a
persistent theme regarding cadmium contamination of soils has been the claim that
phosphate fertilizers, especially `triple super phosphate' (TSP) directly adds Cd
 to the soil due to high levels of contamination.
 On the other hand,
cite{JayasumanaSP15} reported a maximum  of $\sim$ 38
 mg of As per kg of phosphate fertilizer but not cadmium,
 in  some fertilizers imported to  Sri Lanka.
While the build up in the soil of Cd and fluoride contamination from fertilizers is
invoked by many writers. e.g., ~\cite{Premarathne2006,Loganathan2008,TothHeavyMet-MALs2016}, simple
 calculations
 suggest that cadmium or arsenic build up in the NCP in Sri Lanka (or anywhere else)
due to fertilizer use
 is a very unlikely scenario. 
%
%

If we assume that 25 kg of phosphate fertilizer per hectare are applied into a
 depth of 15 cm in
 paddy cultivation, a soil volume
of 1.5$\times10^6$ liters is treated; 
 hence we assume a dry soil weight of $ \simeq 10^6 $ kg.
 An extremely contaminated sample of TSP may contain
 50 mg of Cd, and 50 mg of As per kg of fertilizer. Even such a contaminated fertilizer
will add only (50x25 mg)/(10$^6 $) = 1.25 $\mu$g of Cd and As per kilo of soil per
application, while typical threshold values are  5 mg of As and  3 mg of Cd per kg of soil.
~
Hence, even if {\it all} the Cd and As remain in the soil, with no wash-off and continue
to be bio-available, it will take 4000 applications to reach toxic-soil thresholds of
As, and 2400 applications to reach the threshold for Cd. Even at two applications per year,
and with uninterrupted use of a heavily contaminated TSP containing 50 mg/kg of Cd or As,
 it will take 1.2 millenia  to
build up the Cd level in soil to dangerous thresholds, and 2 millenia for
 As. In reality much of the Cd or As from the TSP
applied to the soil become part of runoff during monsoons, and also converts to 
bio-unavailable forms held in the soil. Thus the time needed for Cd impurities
in fertilizers to reach thrshold levels becomes  longer than even millenia time scales.
 Hence the use of
``contaminated fertilizer'' cannot be considered the origin of higher levels of
cadmium found in agricultural soils.

The Cd and As in the NCP soils of Sri Lanka are 
 mostly likely to be from normal geological sources and from
 acid rain linked to coal-fired
 power stations that are abundant along the South Indian coastline.
 A small contribution
 comes from  vehicles and tractors using fossil fuels. 
The heavy rainfall
 in the wet zone  results in prolonged submerging of paddy soils producing
 extreme reducing conditions~\citep{Chandrajith2005b}. 
This facilitates the mobilization of Cd and the
subsequent Cd uptake  by the plant. In  the dry zone,
soils are irrigated avoiding prolonged submerging. A major 
 source can be from the phyto-accumulation of cadmium and other metal toxins
by leafy vegetation, crops,
 paddy and grasses~\citep{CdPlants2013,McWilliams2009}.
%
 When such plant matter is recycled to the soil through normal decomposition or
 as compost fertilizer, the Cd and similar toxins
accumulated in the plants are returned
 to the soil at an  increased concentration (factor of 30-100). This may
explain why agricultural soils contain more cadmium than some virgin soils. 
Also, use of industrial sludge,  dredging from tanks (``weva''), or composted
urban waste or waste from livestock farms as components in fertilizer
 can contribute
 unknown amounts of metal toxins to
agricultural soils.
\section{The dietary load of Cd and other toxins}
\label{diet.sec}
The existence of  interactions among commonly found ions like Mg$^{2+}$, and
toxic ions like F$^-$, Cd$^{2+}$, implies that the accepted MALs for such toxic
ions have to be reviewed when used for regulatory purposes.
A review of the  dietary load of such common toxins in the presence of
other salts and micro-nutrients is hence needed.

In the case of fluoride, while a lower beneficial limit  is  desired to
prevent dental carries, the American Dental Association also provides a
daily tolerable upper limit of 10 mg  {\it irrespective of the body weight}
of fluoride intake for adults from all sources, beyond which it
becomes a health risk~\citep{ADA-Fluo}. The Environmental Protection agency
sets the limit at
6 mg/day~\citep{EPA2011}.
Considering the exceptionally long half-life of cadmium (in the body)
 and that daily or
weekly  ingestion in food would have a negligible effect on overall
exposure, a provisional tolerable monthly intake (PTMI)  of 25 $\mu$g of Cd
per kg of body weight has been established~\citep{JECFA2011}. For 
convenience we use the equivalent provisional tolerable daily intake (PTDI) of 0.833
$\mu$g of Cd per kg of body weight.  

%
%
This PTMI makes no mention of how to allow for the presence
of other ions derived from, e.g.,  Zn, Se, Mg, and F. Studies by
~\cite{Diyaba-Rice-2016}  provide data (Table~\ref{rice-tab}) on
 the concentration
of As, Cd, Se, as well as Zn found in rice grown in three climatic zones, namely, the
 Dry Zone (DZ), the intermediate zone (IZ), and the Wet Zone (WZ). The
endemic areas are in the DZ, while the WZ is essentially free of the disease. 
\begin{table}
\caption{Metal content in rice for the three climnate zones. Amounts in $\mu$g/kg.  Where
possible median amounts have been used from~\cite{Diyaba-Rice-2016}. The fluoride
and aluminum values are estimated~\citep{TegegneEthi2013,Amara-Dharmaguna2014} and no
 distinction between zones is made.}
\center
\begin{tabular}{l l l l l }
\hline\\
Rice              &  unit     &       DZ       &  IZ       &      WZ  \\
As                & ppb       &     29    &  34       &     40    \\
Cd                & ppb       &     52    &     38    &     79    \\
Se                & ppb       &     26    &     14    &     19    \\
Zn                & ppm       &     14    &  14       &     16    \\
F$^{a}$           & ppm       &    6.0    &  6.0      &   6.0     \\
Al$^a$            & ppm       &  204      &  204      &   204     \\
\hline  
$^a$ estimated
\end{tabular}
\label{rice-tab}
\end{table}
%
The earlier data for As and Cd in rice by \cite{Meharg2009,Meharg2013} are consistent
 with the more recent data.

 The most striking feature of these data is the presence of significantly higher
amounts of As and Cd in the rice consumed by people living
in the wet zone where there is no CKDu. On the average, 50\% more cadmium
 and 40\% more As are
 ingested via  rice by residents of a typical WZ-village, e.g, Bandaragama.
 We have chosen Bandaragama since  data for the
heavy metal concentrations in the vegetables  in the
 area have been reported~\citep{Premarathne2006}. They are given in Table~\ref{veg-tab}.
 The same concentration trends
for Cd, As and Zn found in rice for the DZ, IZ, and WZ may be expected
for the leafy vegetables as well.
\begin{table}
\caption{Concentration (mg/kg) of some heavy metals in Bandaragama, a WZ village
$\sim$35 km SSE of Colombo, Sri Lanka~\citep{Premarathne2006}}
\center
\begin{tabular}{p{2.5 cm}p{1.0 cm}p{1.0 cm}p{1.0 cm}p{1.0 cm}}
\hline\\
crop                  &  Cd    &  Cu   &   Pb   & Zn  \\
Ipomea aquatica (Kangkung)      &  0.37  &  12.2 &   9.24 & 56.6 \\
Centella asiatica (Gotukola)     &  0.54  &  6.03 &   8.75 & 48.1 \\  
Alternanthera Sesilis (Mukunu-venna)  &  0.17  &  8,6  &   10.4 & 72.6 \\
\hline
\end{tabular}
\label{veg-tab}
\end{table}
A typical Sri Lankan diet includes rice, vegetables, lentils, and some fish or meat.
 Lentils
do not bio-accumulate metal toxins to any significant extent, and the meat and fish
 components are quite small in comparison to Western diets.
Hence the main source of metal toxins is from rice and vegetables. 
The total intake from
rice alone has been estimated in Table 2 of ~\cite{Diyaba-Rice-2016},  where
values of 0.6 $\mu$g and 0.32 $\mu$g  of Cd per kg of body weight (kgBW) from rice
for the WZ and DZ respectively are given. Approximately 20-30\% more are contributed from
vegetables, tea, and other food items, boosting the values to 0.8  and 0.43 
$\mu$g  of Cd per kgBW, for the WZ and DZ respectively.
 This total dietary intake, where we have {\it not} allowed for bioavailability
etc,  for the WZ is at
the threshold of 0.833  $\mu$g  of Cd per kgBW (i.e, the PTDI of Cd) stipulated by
the WHO~\citep{JECFA2011}. The intake in the DZ is considerably below the PTDI.
 Thus, judging by these considerations alone, the DZ-resident
 has not been {\it not under threat} of Cd toxicity, while the WZ-resident
 has  been  under threat or at least under a hazard.
 However, no cadmium toxicity in
any form is seen in the wet zone, even though the rural WZ resident consumes the
traditional ``rice and curry'' diet! 

There are several  simple factors that resolve the above puzzle.
 They are (i) co-action of Zn
and Se which are in concentrations exceeding those of As and Cd by a factor of the
order of 500 (for Zn), as seen in table~\ref{rice-tab}; (ii) only about 30\% of the cadmium is in
a bio-available (i.e., `exchangeable') form, and (iii) less than 5\% of the ingested heavy metals are  absorbed by the gut.

It is known that Se and Zn in the diet have an antagonistic action on
 Cd~\citep{ARLS-review,BrzoskaCdZn2001,MatovicCdZnMg2011}. The behaviour
of these ions in the diet is not likely to be related
to the ion-pairing process described previously, applicable to drinking water.
The European union allows the sale of oysters with high amounts of Cd, if Zn is 
also present in significant quantities to compensate the action of Cd. The rice
consumed by the residents of Sri Lanka may contain Cd, but also contains overwhelming
amounts of Zn ions.  Cd has a less favorable Gibbs free energy for ion association
compared to Zn, and Zn is chemically and  bio-chemically more active than Cd.
 While the WHO study noted a deficiency of Se in CKDu patients,
this is more likely to be an issue of late stage CKDu patients rather than a
reflection of their diet, as the rice has selenium in significant amounts.

The PTDI thresholds are stipulated for the action of, say, Cd ions when they are
assumed to be fully bioavailable and able to act independently. If the 
bioavailable Cd and Zn  concentrations are denoted by
 $\left[\mbox{Cd}^{2+}\right]$ and $\left[\mbox{Zn}^{2+}\right]$, assuming that cadmium ions do not
act until all the zinc ions are depleted, then we define an effective cadmium concentration 
$\left[\mbox{Cd}^{2+}_{eff}\right]$=$\left[\mbox{Cd}^{2+}\right]$-$\left[\mbox{Zn}^{2+}\right]$. 
The PTDI should be compared with daily intake calculated from 
 $\left[\mbox{Cd}^{2+}_{eff}\right]$ and not
from the uncorrected amount. 

\section{Conclusion}
We have examined the toxicity of Cd, F, As and other chemical agents in
 their ionic  form and their
 interactions in drinking water, as well as their presence in the diet. The toxicity
enhancement or suppression in water is modeled via the formation of strategic ion pairs. The 
electro-chemistry of anion-cation interactions in hard drinking water containing
Mg$^{2+}$ ions is able to completely account for the experimental observations
of renal damage observed in experiments on mice by ~\cite{SynergyBandara2017}. In particular,
 interactions between fluoride
 and magnesium, and also fluoride and cadmium cause enhanced nephrotoxicity. Strong interactions between
 fluoride and aluminum ions, or fluoride and zinc ions are
found. No enhancement of
arsenite toxicity due to hardness is predicted from this electrochemical analysis. The results  
strongly suggest that stagnant well water consumed by some residents in the 
CKDU-affected areas, and containing F$^-$, Cd$^{2+}$ and Mg$^{2+}$ can be a definitive cause of the disease. Calcium hardness is found to be irrelevant within our toxicity
 model of ion association, as the Gibbs free energy of association with calcium ions with fluoride,
  arsenate or arsenite is positive (i.e, not favoured). 

The data for the toxicity of heavy metals in the diet underlines the importance of
 micro-nutrient ions like Zn and Se that can play a role in suppressing the toxicity of
 Cd, F and possibly also arsenic. While the uncorrected Cd and As dietary load of {\it wet zone}
 residents
 in Sri Lanka is close to or above the WHO safety threshold, they become very safe when
 the effect of Zn, Se, and low bio-availability are taken into account. The Dry Zone residents
 do not have  heavy Cd and As loads in their diet, even when we use uncorrected dietary 
loads to compare with
 the PTMI values of the WHO.
It is suggested that the use of unmodified safety thresholds (e.g., MALs, PTDI etc.)
for the toxicity of ions is valid only for single-ions in solution, or when there is a very large
number of ions  of similar electrochemical activity where a multiplicity of competing interactions restores the simple MAL model when the central-limit theorem begins to hold sway.

%

\begin{thebibliography}{64}
\providecommand{\natexlab}[1]{#1}
\providecommand{\url}[1]{{#1}}
\providecommand{\urlprefix}{URL }
\expandafter\ifx\csname urlstyle\endcsname\relax
  \providecommand{\doi}[1]{DOI~\discretionary{}{}{}#1}\else
  \providecommand{\doi}{DOI~\discretionary{}{}{}\begingroup
  \urlstyle{rm}\Url}\fi
\providecommand{\eprint}[2][]{\url{#2}}

\bibitem[{{Agriculture Dept.}(2013)}]{Agri-dept}
{Agriculture Dept} Sl (2013) crop recommendations. Website:
http://www.agridept.gov.lk/index.php/si/
crop-recommendations/903

\bibitem[{Amarasooriya and Dharmagunawardhane(2014)}]{Amara-Dharmaguna2014}
Amarasooriya AAGD, Dharmagunawardhane HA (2014) Leaching of aluminum and its
  incorporation to rice during cooking under different fluoride concentrations
  in water. SAITM Research Symposium on Engineering Advancements, Malabe, Sri
  Lanka., website:http://www.saitm.edu.lk/fac\_of\_eng/RSEA/
SAITM\_RSEA\_2014/imagenesweb/12.pdf

\bibitem[{{American Dental Association}(2005)}]{ADA-Fluo}
{American Dental Association} (2005) Fluoridation facts. Tech. rep., website:
  www.ada.org/\~/media/ADA/Member\%20Center
FIles/fluoridation\_facts.ashx

\bibitem[{ARL(2012)}]{ARLS-review}
ARL (2012) Tech. rep., Analytical Research labs, Inc., Phoenix, Arizona, USA,
  website:http://www.arltma.com/Articles/
CadmiumToxDoc.htm

\bibitem[{{ASTDR, USA}(2013)}]{ATSDR2013}
{ASTDR, USA} (2013) {C}admium toxicity what are the {U}.{S}. standards for
  cadmium exposure? Website: https://www.atsdr.cdc.gov/csem/csem.asp?csem=6
\&po=7

\bibitem[{Ben-Naim(1992)}]{Ben-Naim1992}
Ben-Naim A (1992) Statistical thermodynamics for chemists and biochemists.
  Plenum Press. New York, USA

\bibitem[{Brz\'{o}ska et al (2001)}]{BrzoskaCdZn2001}
Brz\'{o}ska MM, Moniuszko-Jkoniuk J (2001) Interactions between cadmium and
  zinc in the orgnism. Food and Chemical Toxicology 39

\bibitem[{Bustamante and Feraille(2007)}]{Na-Pump2007}
Bustamante M, Feraille E (2007) Sodium in Health and Disease. CRC press

\bibitem[{Chandrajith et~al(2005)Chandrajith, Dissanayake, and
  Tobschall}]{Chandrajith2005b}
Chandrajith R, Dissanayake CB, Tobschall HJ (2005) Geochemistry of trace
  elements in paddy (rice) soils of sri lanka- simplications for iodine
  deficiency disorders. Geochem Envron \& Health 27:55--64

\bibitem[{Collins(1997)}]{Collins1997}
Collins KD (1997) Biophys J 72:65--76

\bibitem[{Dahanayake et~al(2012)Dahanayake, Wijewardahana, Jayasumana, and
  Paranagama}]{Paranagama2012}
Dahanayake KS, Wijewardahana KMRC, Jayasumana MACS, Paranagama P (2012)
  Presence of high levels of arsenic in internal organs of deceased patients
  with chronic kidney disease of unknown aetiology (ckdu): three case reports.
  In: Proceedings of the Research Symposium on Chronic Kidney Disease of
  Unknown Aetiology (CKDu), Sri Lanka Medical Association, Colombo, Sri Lanka.,
  website:https://issuu.com/slmanews/
docs/ckdu\_abstract\_book.html

\bibitem[{Dharma-wardana (2015)Dharma-wardana, Amarasiri, Dharmawardene,
  and Panabokke}]{Dharma-wardana2015}
Dharma-wardana MWC, Amarasiri S, Dharmawardene N, Panabokke CR (2015) Chronic
  kidney disease of unknown aetiology and ground-water ionicity: study based on
  sri lanka. Environ Geochem Health 37:221--231

\bibitem[{Dissanayake(2005)}]{Dissanayake2005}
Dissanayake CB (2005) Water quality in the dry zone of sri lanka. some
  interesting health aspects. Journal of National Science Foundation of Sri
  Lanka 33:161--168

\bibitem[{Diyabalanage et~al(2016{\natexlab{a}})Diyabalanage, Abekoon,
  Watanabe, and {\it et al.}}]{Diyabalanage2016}
Diyabalanage S, Abekoon S, Watanabe I, {\it et al} (2016{\natexlab{a}}) Has
  irrigated water from mahaweli river contributed to the kidney disease of
  uncertain etiology in the dry zone of sri lanka? Environ Geochem Health
  38:439--454, \doi{10.1007/s10653-015-9749-1}

\bibitem[{Diyabalanage et~al(2016{\natexlab{b}})Diyabalanage, Navarathna,
  Abeysundara, and {\it et al.}}]{Diyaba-Rice-2016}
Diyabalanage S, Navarathna T, Abeysundara TA, {\it et al} (2016{\natexlab{b}})
  Trace elements in native and improved paddy rice from different climatic
  regions of sri lanka: implications for public health. Springer Plus 5:1684,
  \doi{10.1186/s40064-016-3547-9}

\bibitem[{EPA(2011)}]{EPA2011}
EPA (2011) S-environmental protection agency, drinking water regulations on
  fluoride. Tech. rep., Environmental Protection Agency, USA, website:
  https://safewater.zendesk.com/hc/en-us/
articles/212076577-4-What-are-EPA-s-drinking-water
 -regulations-for-fluoride-

\bibitem[{Fox(2011)}]{Reuters2011}
Fox M (2011) {U}.{S}. lowers limits for fluoride in water. Website:
  http://www.reuters.com/article/
us-usa-fluoride-idUSTRE7064CM20110108

\bibitem[{Gracia-Tabanino et~al(2005)Gracia-Tabanino, J., and
  Oliver}]{Trabanino05}
Gracia-Tabanino R, J D, Oliver JA (2005) Proteinuria and chronic renal failure
  in the coast of {E}l {S}alvador: detection with low cost methods and
  associated factors. Nefrologia XXXV(1)

\bibitem[{Illeperuma et~al(2009)Illeperuma, Dharmagunawardhane, and
  Herath}]{IllepAl}
Illeperuma OA, Dharmagunawardhane HA, Herath KRP (2009) Dissolution of
  aluminium from substandard utensils under high fluoride stress: A possible
  risk factor for chronic renal failures in the {N}orth-{C}entral province.
  Journal of the National Science Foundation of Sri Lanka 37:219--222

\bibitem[{ITFG(2014)}]{TaskForceGly}
ITFG (2014) Glyphosate and chronic kidney disease - {S}ri {L}anka. Tech. rep.,
  Industry Task Force on Glyphosate,
  http://www.glyphosate.eu/glyphosate-
and-chronic-kidney- disease-sri-lanka

\bibitem[{Jayasekara et~al(2013)Jayasekara, Dissanayake1, Adhikari, and
  Bandara}]{Jayasekera2012}
Jayasekara JMKB, Dissanayake1 DM, Adhikari SB, Bandara P (2013) Geographical
  distribution of chronic kidney disease of unknown origin in north central
  region of {S}ri {L}anka. Cey Medical J 58:6--9

\bibitem[{Jayasinghe et~al(2015)Jayasinghe, Herath, and
  Wickremasinghe}]{Jayasinghe-RO-2015}
Jayasinghe P, Herath B, Wickremasinghe N (2015) {T}echnical review report based
  on visit to {A}nuradhapura ckdu affected areas; review of input-output water
  of reverse-osmosis installtions. Tech. rep., COSTI (Coordinating Office for
  Science and Technology Innovation, Sri Lanka), website:
  https://dh-web.org/placenames/
posts/COSTI-Jaysinghe-RO.pdf

\bibitem[{Jayasumana (2011)Jayasumana, Parangama, and
  Amarasinghe}]{Jayasumana11}
Jayasumana C, Parangama P, Amarasinghe M (2011) Chronic kidney disease of
  unknown etiology (ckdu) and~ arsenic in ground water in {S}ri {L}anka.
  presence of arsenic in pesticides used in {S}ri {L}anka. In: Proc. workshop
  on Challenges in groundwater Management in {S}ri {L}anka, website:
  http://www.wrb.gov.lk/web/images/stories/
downloads/Scientific\_Reportsproceeding\_07\_april\_11.pdf

\bibitem[{Jayasumana (2014)Jayasumana, Gunatilake, and
  Senanayake}]{JayasumanaGly2015}
Jayasumana C, Gunatilake S, Senanayake P (2014) Glyphosate, hard water and
  nephrotoxic metals: Are they the culprits behind the epidemic of chronic
  kidney disease of unknown etiology in sri lanka? Int J Environ Res Public
  Health 11:2125--2147

\bibitem[{Jayasumana et~al(2015)Jayasumana, Fonseka, Fernando, Jayalath,
  Amarasinghe, Siribaddana, Gunatilake, and Paranagama}]{JayasumanaSP15}
Jayasumana C, Fonseka S, Fernando A, Jayalath K, Amarasinghe M, Siribaddana S,
  Gunatilake S, Paranagama P (2015) Phosphate fertilizer is a main source of
  arsenic in areas affected with chronic kidney of unknown etiology in sri
  lanka. Springer Plus 4:90, \doi{10.1186/s40064-015-0868-z.}

\bibitem[{Jayasumana et~al(2013)Jayasumana, Paranagama, Amarasinghe,
  Wijewardane, Dahanayake, Fonseka, Rajakaruna, Mahamithawa, Samarasinghe, and
  Senanayake}]{jayasumana13}
Jayasumana MACS, Paranagama PA, Amarasinghe MD, Wijewardane KMRC, Dahanayake
  KS, Fonseka SI, Rajakaruna KDLM, Mahamithawa AMP, Samarasinghe UD, Senanayake
  VK (2013) Possible link of chronic arsenic toxicity with chronic kidney
  disease of unknown etiology in {S}ri {L}anka. Journal of Natural Science
  Research 3(1):64--73

\bibitem[{Jayatilake (2013)Jayatilake, S., Maheepala, Metha, R., and
  project Team.}]{WHO2}
Jayatilake N, S M, Maheepala P, Metha, R F, project Team CNR (2013) Chronic
  kidney disease of uncertain aetiology, prevalence and causative factors in a
  developing country. BMC Nephrology 14:180

\bibitem[{JECFA(2011)}]{JECFA2011}
JECFA (2011) Joint {FAO/WHO} food standards programme, {CODEX} committee on
  contaminants in foods fifth session. Tech. rep., WHO-FAO, Joint FAO/WHO
  Expert Committee on Food Additives (JECFA), website:
  ftp://ftp.fao.org/codex/meetings/CCCF/cccf5/
cf05\_INF.pdf

\bibitem[{Jessani et~al(2014)Jessani, Bux, and Jafar}]{ckdu-Pakistan}
Jessani S, Bux R, Jafar T (2014) Prevalence, determinants, and management of
  chronic kidney disease in {K}arachi, {P}akistan - a community based
  cross-sectional study. BMC Nephrology 15:90

\bibitem[{Jin (2011)Jin, Qian, Lu, Yang, and Bi}]{XiaJinDFT-AlF2011}
Jin X, Qian Z, Lu B, Yang W, Bi S (2011) Density functional theory study on
  aqueous aluminum¿fluoride complexes: Exploration of the intrinsic
  relationship between water-exchange rate constants and structural parameters
  for monomer aluminum complexes. Environmental Science \& Technology
  45(1):288--293, \doi{10.1021/es102872h},
  \urlprefix\url{http://dx.doi.org/10.1021/es102872h}, pMID: 21133367,
  \eprint{http://dx.doi.org/10.1021/es102872h}

\bibitem[{Kerslake et~al(1946)Kerslake, Schmitt, and Thomas}]{KerslakeFeF1946}
Kerslake J, Schmitt H, Thomas N (1946) Iron salts as coagulants. J Am Water
  Works Assoc 38:1161--1169

\bibitem[{Koneshan (1998)Koneshan, Rasaiah, Lynden-Bell, , and
  Lee}]{Rasiah1998}
Koneshan S, Rasaiah JC, Lynden-Bell RM, , Lee SH (1998) Solvent structure,
  dynamics, and ion mobility in aqueous solutions at 25 °c. The Journal of
  Physical Chemistry B 102:4193--4204, \doi{10.1021/jp980642x}

\bibitem[{Kumar(2017)}]{VijayaKumarPrCom2017}
Kumar V (2017) private communication

\bibitem[{Levine et~al(2016)Levine, Redmon, Elledge, Wanigasuriya, and {\it et
  al}.}]{LevineWanig2016}
Levine KE, Redmon JH, Elledge MF, Wanigasuriya KP, {\it et al} (2016) Quest to
  indentify geochemical risk factors associated with chronic kidney disease of
  unknown etiology (ckdu) in an endemic region of sri lanka - a multimedia
  laboratory analysis of biological, food, and environmental samples. Environ
  Monit Assess 188

\bibitem[{Lin et~al(2014)Lin, Shao, Luo, Ou-yang, Zhou, Du, He, Wu, Xu, and
  Chen}]{BoLin14}
Lin B, Shao L, Luo Q, Ou-yang L, Zhou F, Du B, He Q, Wu J, Xu N, Chen J (2014)
  Prevalence of chronic kidney disease and its association with metabolic
  diseases: a cross-sectional survey in {Z}hejiang province, eastern {C}hina.
  BMC Nephrology 15

\bibitem[{Loganathan et~al(2008)Loganathan, Headly, and Grace}]{Loganathan2008}
Loganathan P, Headly MJ, Grace ND (2008) Pasture soils contaminated with
  fertilizer-derived cadmium and fluorine. Rev Environ Contam Toxicol
  129:29--66

\bibitem[{Manthrithilake(2016)}]{Manthrithilake2016}
Manthrithilake H (2016) Ckdu: Are we shooting the right target?
 http://dh-web.org/place.names/posts/
Manthithilake-CKDu.pdf  accessed 12.Nov.2016

\bibitem[{Marcus(1991)}]{Marcus87}
Marcus I (1991) Part 5- gibbs free energy of hydration at 298.15 k. J Chem Soc
  Faraday Tranas 87:2995--2999

\bibitem[{Matovi\'{c} et~al(2011)Matovi\'{c}, Buha, Bulat, and
  Duki\'{c}-\'{C}osi\'{c}}]{MatovicCdZnMg2011}
Matovi\'{c} V, Buha A, Bulat Z, Duki\'{c}-\'{C}osi\'{c} D (2011) Cadmium
  toxicity revisited: Focus on oxidative stress induction and interactions
  with, zn and mg. Arh Hig Rada Toksikol 62:65--76

\bibitem[{McWilliams(2009)}]{McWilliams2009}
McWilliams JE (2009) Just Food: Where Locavores Get It Wrong and How We Can
  Truly Eat Responsibly. Little, Brown and Co., New York. USA

\bibitem[{Meharg et~al(2009)Meharg, Williams, Adomako, Y., Deacon, Villada,
  Campbell, Sun, Zhu, and Feldman}]{Meharg2009}
Meharg AA, Williams PN, Adomako E, Y LY, Deacon C, Villada A, Campbell RCJ, Sun
  G, Zhu YG, Feldman J (2009) Geographical variation in total and inorganic
  arsenic content of polished (white) rice. Environ Sci Technol 43:1612--1617

\bibitem[{Meharg et~al(2013)Meharg, Norton, Deacon, Williams, Adomako, Price,
  Zhu, Li, Zhao, and McGarth}]{Meharg2013}
Meharg AA, Norton G, Deacon V, Williams P, Adomako E, Price A, Zhu Y, Li G,
  Zhao F, McGarth S (2013) Variation in rice cadmium related to human exposure.
  Environ Sci Technol 47:5613--5618

\bibitem[{{MINTEQ}(2000)}]{MINTEQ}
{MINTEQ} (2000) {MINTEQ}, version 3 is a freeware chemical equilibrium model
  for calcularting metal speciation, equilibria, sorption etc. for natural
  waters. Website \url{https://vminteq.lwr.kth.se/}

\bibitem[{Nanayakkara et~al(2014)Nanayakkara, Senevirathna, Abeysekera,
  Chandrajith, Ratnatunga, Gunarathne, Yan, Hitomi, Muso, Komiya, Harada, Liu,
  Kobayashi, Okuda, Sawatari, Matsuda, Yamada, Watanabe, Miyataka, Himeno, and
  Koizumi}]{NanayakkaraS14}
Nanayakkara S, Senevirathna S, Abeysekera T, Chandrajith R, Ratnatunga N,
  Gunarathne E, Yan J, Hitomi T, Muso E, Komiya T, Harada KH, Liu W, Kobayashi
  H, Okuda H, Sawatari H, Matsuda F, Yamada R, Watanabe T, Miyataka H, Himeno
  S, Koizumi A (2014) An integrative study of the genetic, social and
  environmental determinants of chronic kidney disease characterized by
  tubulointerstitial damages in the north central region of {S}ri {L}anka.
  Journal of Occupational Health 56:28--38

\bibitem[{Panabokke(2007)}]{Pana1}
Panabokke CR (2007) Groundwater conditions in {S}ri {L}anka. National Science
  Foundation of Sri Lanka, Colombo. Sri Lanka

\bibitem[{Premarathne(2006)}]{Premarathne2006}
Premarathne HMPL (2006) Soil and crop contamination by toxic trace elements.
  Master's thesis, Post Graduate Institute of Agriculture, University of
  Peradeniya, Sri Lanka.

\bibitem[{Premarathne et~al(2011)Premarathne, Hettiarachchi, and
  Indrarathne}]{Premarathne2011}
Premarathne HMPL, Hettiarachchi GM, Indrarathne SP (2011) Toxic trace metal
  concentration in crops and soils colected from intensively cultivated areas
  of sri lanka. Pedologist 54:230--240

\bibitem[{Reddy et al (2013)}]{DVReddy2013}
Reddy DV, Gunasekar A (2013) Chronic kidney disease in two coastal districts of
  andhra pradesh, india: role of drinking water. Environ Geochem Health
  35:439--454, \doi{10.1007/s10653-012-9506-7}

\bibitem[{Salis and Ninham(2014)}]{SalisNinham2014}
Salis A, Ninham BW (2014) Models and mechanisms of hofmeister effects in
  electrolyte solutions, and colloid and protein systems revisited. Chem Soc
  Rev 43:7358--7377

\bibitem[{Shannon(1976)}]{Shannon1976}
Shannon RD (1976) Revised effective ionic radii and systematic studies of
  interatomic distances; electronic table of shannon ionic radii, {J}. {D}avid
  {V}an {H}orn, 2001. Acta Cryst A32:751--767, website:
  http://v.web.umkc.edu/vanhornj/shannonradii.htm

\bibitem[{Smith(2004)}]{Smith2004}
Smith PE (2004) Cosolvent interactions with biomolecules: Relating computer
  simulation data to experimental thermodynamic data. Journal of Physical
  Chemistry B 108:18,716--18,724

\bibitem[{Smith and Raymond(1988)}]{SmithRay88}
Smith PH, Raymond KN (1988) Inorg Chem 27:1056--1061

\bibitem[{Stritsis and Claassen(2013)}]{CdPlants2013}
Stritsis C, Claassen N (2013) Cadmium uptake kinetics and plants factors of
  shoot cd concentration. Plant soil 367:591--603

\bibitem[{Tegegne et~al(2013)Tegegne, Chandravanshi, and
  Zewge}]{TegegneEthi2013}
Tegegne B, Chandravanshi SB, Zewge F (2013) Bull Chem Soc Ethiop 27:179--189

\bibitem[{Thammitiyiagodage(2012)}]{ThammitiyaNCPwater2012}
Thammitiyagodage MG (2012) Identification of possible causative factors in
  chronic kidney disease(ckd) in north central province (ncp) by an animal
  experimentation. Master's thesis, University of Peradeniya, Sri Lanka, mphil
  Thesis

\bibitem[{Tomljenovic and Shaw(2012)}]{Al-Shaw2011}
Tomljenovic L, Shaw C (2012) Aluminum vaccine adjuvants: Are they safe? Current
  Medicinal Chemistry 17:2630--2637

\bibitem[{T\`{o}th et~al(2016)T\`{o}th, Hermann, Da~Silva, and
  Montanarella}]{TothHeavyMet-MALs2016}
T\`{o}th G, Hermann T, Da~Silva MR, Montanarella L (2016) Heavy metals in
  agricultural soils of the european union with implica- tions for food safety.
  Environment International 88:299--239

\bibitem[{Wanigasuriya et~al(2011)Wanigasuriya, Peiris-John, and
  Wickremasinghe}]{KamaniCd11}
Wanigasuriya KP, Peiris-John RJ, Wickremasinghe R (2011) Chronic kidney disease
  of unknown aetiology in {S}ri {L}anka: is cadmium a likely cause? BMC
  Nephrology 12:32, website:
  http://www.biomedcentral.com/1471-2369/12/32, downloaded 21-Nov-2013

\bibitem[{Wasana et~al(2016)Wasana, Aluthpatabendi, Kularatne, Wijekoon,
  Weerasooriya, and Bandara}]{WasanaWaterQ-CdF2016}
Wasana HMS, Aluthpatabendi D, Kularatne WMTD, Wijekoon P, Weerasooriya R,
  Bandara J (2016) Drinking water quality and chronic kidney disease of unknown
  etiology (ckdu): synergic effects of fluoride, cadmium and hardness of water.
  Environmental Geochemistry and Health 38(1):157--168,
  \doi{10.1007/s10653-015-9699-7},
  \urlprefix\url{http://dx.doi.org/10.1007/s10653-015-9699-7}

\bibitem[{Wasana et~al(2017)Wasana, Perera, Panduka De S.~Gunawardena,
  Fernando, and Bandara}]{SynergyBandara2017}
Wasana HMS, Perera GDRK, Panduka De S~Gunawardena PdS, Fernando PS, Bandara J
  (2017) {WHO} water quality standards vs synergic effect(s) of fluoride, heavy
  metals and hardness in drinking water on kidney tissues. Nature-scientific
  reports \doi{10.1038/srep42516}

\bibitem[{Weave et~al(2015)Weave, Fadrowski, and Jaar}]{GlobalDimension2015}
Weave VM, Fadrowski GF, Jaar BG (2015) Global dimensions of chronic kidney
  disease of unknown etiology (ckdu): a modern era environmental and/or
  occupational nephropathy? BMC Nephrology 16(1):1,
  \doi{10.1186/s12882-015-0105-6}

\bibitem[{{WHO-SL-reports}(2013)}]{WHO1}
{WHO-SL-reports} (2013) Chronic kidney disease of unknown aetiology ({CKD}u): a
  new threat to health.
 http://dh-web.org/place.names/posts/index.html\#ckdu accessed 12.
  Nov.2013

\bibitem[{World-Bank(2016)}]{WorldBankFert}
World-Bank (2016) Fertilizer consumption (kilograms per hectare of arable
  land). Tech. rep., Food and Agriculture Organization, website:
  http://data.worldbank.org/indicator/AG.CON.
FERT.ZS

\bibitem[{Yeo et~al(2012)Yeo, Tan, Koh, Khan, Nilar, and Go}]{Nilar2012}
Yeo WK, Tan KL, Koh SB, Khan M, Nilar S, Go ML (2012) Exploration and
  optimization of structure-activity relationships in drug design using the
  taguchi method. Chem{M}ed{C}hem 7:977--982

\end{thebibliography}

\end{document}